# Domain Wall Resistance in Perpendicular (Ga,Mn)As: dependence on pinning

K. Y. Wang[1,2,*], K. W. Edmonds[3], A. C. Irvine[4], J. Wunderlich[2], K. Olejnik[2], A. W. Rushforth[3], R. P. Campion[3], D. A. Williams[2], C. T. Foxon[3], and B. L. Gallagher[3]

[1] SKLSM, Institute of Semiconductors, Chinese Academy of Sciences, Beijing P. O. Box 912, 100083, P. R. China

[2] Hitachi Cambridge Laboratory, Cambridge CB3 0HE, United Kingdom

[3] School of Physics and Astronomy, University of Nottingham, Nottingham NG7 2RD, United Kingdom

[4] Microelectronics Research Centre, Cavendish Laboratory, University of Cambridge, CB3 0HE, United Kingdom

**Abstract:** We have investigated the domain wall resistance for two types of domain walls in a (Ga,Mn)As Hall bar with perpendicular magnetization. A sizeable positive intrinsic DWR is inferred for domain walls that are pinned at an etching step, which is quite consistent with earlier observations. However, much lower intrinsic domain wall resistance is obtained when domain walls are formed by pinning lines in unetched material. This indicates that the spin transport across a domain wall is strongly influenced by the nature of the pinning.

## 1. Introduction

The interplay between magnetization and electrical properties in ferromagnetic systems is central to magnetic storage and sensing applications. The technological value of this subject has led to renewed interest in the properties of magnetic domain walls, including the development of schemes for controlling and manipulating domain walls in memory devices. The details of the interaction between a spin-polarized electrical current and the non-uniform local magnetization at a domain wall is still the subject of considerable debate. Determination of the magnitude and the sign of the electrical resistance associated with a domain wall is crucial for understanding the spin transport and dynamics of domain walls, and for making use of them in spintronic devices.

Studies of domain walls in ferromagnetic semiconductors have proved fruitful, yielding observations of much higher domain wall resistance (DWR) [1,2], and much lower critical currents for domain wall motion [3,4], compared to ferromagnetic metal films. For (Ga,Mn)As/GaAs with in-plane magnetization, the DWR was reported to be negative in sign, and comparable in magnitude to the bulk resistivity [1]. However, it was shown that this can be fully attributed to an extrinsic effect, originating from current redistribution in the vicinity of the domain wall due to differing sign of the planar Hall resistance on either side of the wall [5]. The extrinsic effect was accounted for in a study of DWR in (Ga,Mn)As/(In,Ga)As with perpendicular-to-plane magnetization, by performing measurements on a number of samples with varying current channel width [2]. This study yielded a positive intrinsic domain wall resistance-area product of around 0.5 $\Omega\mu m^2$, which is one to two orders of magnitude larger than the DWR predicted for (Ga,Mn)As using a Landauer-Büttiker formalism [6,7]. The observed large positive DWR was therefore attributed to a disorder-induced mixing of spin channels in the ferromagnetic semiconductor [2], and compared to the spin-dependent impurity scattering theory of Levy and Zhang [8].

---

[*] E-mail: kywang@semi.ac.cn

In this letter, we compare the DWR for domain walls in a perpendicularly magnetized (Ga,Mn)As/(In,Ga)As layer for two different pinning centers: a physical step produced by partial etching of the layer (similar to Ref. 2), and a line defect resulting from strain-relaxation in the (In,Ga)As buffer layer. The latter have been shown to act as strong pinning potentials for domain walls, inhibiting their motion along the [110] orientation which is perpendicular to the line defect [4,9,10]. We find that the intrinsic domain wall resistance at the etching step is much larger than that at the nonetched pinning line. The DWR observed for walls pinned at the line defects is in good agreement with semiclassical calculations.

## 2. Experimental Results and Analysis

The 25 nm thick $Ga_{0.95}Mn_{0.05}As$ thin film was grown on a semi-insulating GaAs (001) substrate by Molecular Beam Epitaxy using a modified Varian GEN-II system [11]. A 100 nm thick GaAs buffer layer at 580°C, followed by a 580 nm $In_{0.15}Ga_{0.85}As$ layer at 500°C, were deposited prior to the growth of the (Ga,Mn)As layer at 255 °C. The strain-relaxed (In,Ga)As buffer layer induces a tensile strain in the (Ga,Mn)As film, which results in a perpendicular-to-plane easy magnetic axis [12]. Post-growth annealing was performed in air at 190°C for 24 hours, which is an established procedure for increasing the Curie temperature $T_C$ of (Ga,Mn)As thin films [13]. A 4 µm wide Hall bar structure was fabricated from the (Ga,Mn)As thin film, with 20 µm between neighboring voltage probes, which is shown in Fig. 1a. One end of the Hall bar was wet etched to a depth of 10 to 15nm. We define the *x*-axis as parallel to the current channel, the *y*-axis perpendicular to the current and in-plane, and the *z*-axis as perpendicular to the plane (see Fig. 1(c), inset). Polar magneto-optical Kerr microscopy (PMOKM) and magnetotransport with I =10 µA (Low frequency ac lock-in) have been used to study the domain wall resistance. The $T_C$ of the device is determined to be 122 ± 2 K using PMOKM.

The transport characteristics of the Hall bar device over the studied temperature range are shown in Fig. 1(b) to (d). The sheet resistance $R_S^{unetched}$ of the unetched section of the device is obtained from a measurement of $R_{2a}$ and $R_{2b}$ (see Fig. 1(a)). For the etched section the sheet resistance is obtained as $R_S^{etched} = (2R_1/r - R_S^{unetched})$, where $R_1=(R_{1a}+R_{1b})/2$ and $r=5$ is the ratio of the distance between voltage probes to the channel width. The sheet resistance $R_S$ increases with increasing temperature in both sections (Fig. 1(b)), as is usual for metallic (Ga,Mn)As films below $T_C$. The anomalous Hall angle β= $R_{xy}/R_S$ is obtained for both sections of the device at remanence, after switching the magnetization from up to down, and is shown in Fig. 1(c). The inset to Fig. 1(d) shows the magnetoresistance at 60 K for magnetic field applied along *x, y* and *z* orientations, where the lowest resistance state is obtained when the magnetization is aligned into the channel (*x*) orientation by the applied field. The temperature-dependence of the anisotropic magnetoresistance for magnetization orientations perpendicular to the current channel, $AMR_\perp$ = $(R_{M//y} - R_{M//z})/R_{M//z}$, is shown in Fig. 1(d). The anisotropic magnetoresistance $AMR_{//} = (R_{M//x} - R_{M//z})/R_{M//z}$ has similar magnitude (10% smaller) but with opposite sign across the studied temperature range.

Domain walls are formed at the etching step and at one or more line defects using an external magnetic field. Firstly, a large positive (negative) magnetic field is applied perpendicular to the plane in order to align all of the magnetic domains in the device. Then, the field is swept to certain negative (positive) values, corresponding to the switching magnetic fields of the various magnetic domains, before reducing the external field to zero. PMOKM images of different magnetic configurations of the device, corresponding to one, two or three magnetic domain walls between the voltage probes, are shown in Fig. 2. The domain walls lie perpendicular to the length of bar.

All the domain wall resistance measurements are performed under zero external magnetic field. In order to obtain the domain wall resistance, we measure the longitudinal resistances $R_a$ and $R_b$ for voltage contacts on either side of the device (Fig. 1a), for the magnetic configurations with and without domain walls. $R_a$ and $R_b$ are then averaged in order to remove the contribution from the transverse voltage. The difference in resistance $\Delta R$ between configurations with domain walls present and configurations without magnetic domain walls is shown in Fig. 3a as a function of temperature. The value of $\Delta R$ is found to increase linearly with the number of domain walls pinned between the voltage probes, giving a positive DWR of around 2.1 Ω per wall, which is weakly temperature dependent over the range 30-110 K. As shown in Fig. 3b, the measured DWR for a domain wall pinned at the etching step is much larger, around 6.0 Ω, again varying only slightly over the measured temperature range.

Three contributions to the DWR can be distinguished: (i) an extrinsic contribution due to the polarity change of the anomalous Hall voltage on either side of the wall, causing a current non-uniformity that increases the effective resistance [14]; (ii) a contribution from the anisotropic magnetoresistance due to the in-plane component of the magnetization orientation within the wall [15]; (iii) a contribution due to spin-dependent scattering of electrons traversing the wall. The former two contributions can be obtained from a semiclassical diffusive transport calculation, by numerically solving the current continuity equation $\nabla \cdot J = 0$ and Maxwell's equation $\nabla \times E$ (where $J$ is the current density and $E$ the electric field) for a simulated domain wall structure. The domain wall width and structure was simulated using Landau-Lifschitz-Gilbert micromagnetic calculations [16], with measured anisotropy constant $K \approx 20$ kJ/m$^3$ and typical spin stiffness $A = 0.5$ pJ/m for (Ga,Mn)As [17,18]. A Bloch wall was found to be the most stable configuration, with a width W of around 40nm. Alternatively, the domain wall width can be estimated from W $\sim \pi(A/K)^{1/2}$, with, giving W $\approx$ 20 nm. In the case of the etching step, the most stable configuration occurred for a domain wall positioned just on the etched side of the step. A finite-differences method, similar to that described in Ref. [5], was then applied to calculate the voltage drop across the domain wall, using the simulated wall profile and the measured conductance tensor on either side of the etching step.

Figure 3(c) compares the measured DWR per wall in the unetched part of the Hall bar to the results of the semiclassical calculation with domain wall width W = 0, 20 and 40 nm. The calculation is consistent with analytical expressions for extrinsic DWR [14], and fully reproduces the experimental result, indicating that the intrinsic DWR due to spin-dependent scattering is negligible in the absence of an etching step. The calculated DWR is dominated by the contribution due to the polarity change of the Hall voltage on either side of the wall, which results in a non-uniformity of the current distribution and an increase of the resistance. With increasing wall width, the polarity-change contribution decreases, while the contribution due to AMR within the wall increases linearly, so that the net dependence on the wall width of these contributions to DWR is weak, as shown in Fig. 3(c).

For a domain wall pinned at the etching step, the calculated extrinsic DWR is sensitive to the position of the domain wall relative to the step. Calculated results are shown in Fig. 3(b) for the case that the centre of the domain wall lies slightly on the etched side of the step. This is the most stable position according to the micromagnetic simulation, and is also the position where the extrinsic DWR is maximized. The calculated DWR is again only weakly dependent on the wall width due to the counterbalancing of AMR and Hall voltage contributions. For the whole temperature range studied, the calculated DWR is 2-3 Ω lower than the measured value for domain walls pinned by the etching step. For a domain wall of width 20nm, a 2-3 Ω increase of resistance corresponds to a 15-25% increase of the resistivity in the domain wall region, which is an order of magnitude larger than the anisotropic magnetoresistance contribution.

The intrinsic contribution to DWR can be estimated from the difference between the measured DWR and the semiclassically calculated results. To compare with the measurements of Chiba *et al.* (Ref. 2) on wider Hall bars (25-150μm), the result is expressed as a resistance-area product $R^{int}A = (\Delta R_{meas} - \Delta R_{calc})wt$, with $w$ = 4 μm, and $t$ = 25nm for the non-etched bar and 12.5 ± 2.5 nm for the etching step. Fig. 3d shows $R^{int}A$ versus temperature for the two types of domain walls. For a domain wall trapped at the etching step, $R^{int}A$ is around 0.1-0.15 Ωμm$^2$ over the temperature range studied, which is of the same sign but a factor of 3-4 smaller in magnitude compared to the value obtained by Chiba *et al.* (Ref. 2). The conductivity of the present sample is more than twice as large as in Ref. 2, which may partially account for the reduced $R^{int}A$ according to the Levy-Zhang model, due to reduced scattering [8]. In addition, the higher carrier density in our device compared to Ref. 2 may result in a lower spin-polarization of the current-carrying holes, due to a change in occupancy of the minority- and majority-spin sub-bands [12], further reducing the intrinsic domain wall resistance in our device.

For the domain walls pinned by line defects in the unetched bar, $R^{int}A$ is 0.01 ± 0.02 Ωμm$^2$, *i.e.* zero within the experimental uncertainty. To confirm our results, we also fabricated 8 μm and 20 μm wide devices from another wafer obtained under similar growth conditions. The pinning in these two devices is weaker than that of the studied device. The measured domain wall resistances over the temperature range 30-60K for walls pinned at line defects in the 8μm and 20μm wide devices are (1.7±0.1)Ω and (1.6±0.1)Ω, respectively. Therefore, the DWR is independent of the channel width over this range within the measurement uncertainty. This is expected when the DWR is due to the polarity-change of the Hall voltage, but not for AMR or intrinsic contributions to the DWR. This is therefore consistent with our conclusion that the polarity change contribution is dominant, and the intrinsic DWR is negligible for domain walls pinned at a line defect in our samples.

If the large DWR at the etching step is due to impurity scattering, then the absence of a measurable intrinsic DWR in the unetched case is surprising, as this would be expected to be increased in the disordered region surrounding the line defect. In addition, the stronger pinning at the line defects compared to the etching step should result in narrower domain walls, and hence a larger intrinsic DWR. It may be that during growth or post-growth annealing there is a build-up of paramagnetic impurities (*e.g.*, Mn interstitials) at the pinning line, causing a loss of spin coherence as the current traverses the domain wall, so that the spin-dependent scattering mechanism of Ref. 8 is no longer effective. It worth to note this small value of intrinsic DWR is consistent with calculations performed for (Ga,Mn)As within the Landauer-Büttiker formalism, which predict $R^{int} \approx 10^{-3}$-$10^{-2}$ Ωμm$^2$ under our experimental conditions [7]. Although the mean free path in the (Ga,Mn)As film is of order 1 nm, the Landauer-Büttiker formalism may be still an appropriate approach here due to a relatively long phase coherence length (~ 100 nm at low temperatures) in (Ga,Mn)As [19,20]. Understanding the origin of this dependence, and determining whether it is a universal behavior for ferromagnetic thin films, are important questions for development of the theory of spin transport through non-uniform magnetization.

## 3. Conclusions

In summary, we have investigated the domain wall resistance for two types of domain walls in a (Ga,Mn)As Hall bar with perpendicular magnetization. For domain walls pinned at an etching step, a sizeable intrinsic DWR is inferred which is quite consistent with the observations by Chiba *et al.* [2]. However, much lower intrinsic domain wall resistance is obtained when domain walls are formed by pinning lines in unetched material. This indicates that the spin transport across a domain wall is influenced by the nature of the pinning.

**4.Acknowledgements**

We thank T. Dietl for valuable discussions. This project was supported by EU Grant IST-015728, FP7 Grant no. 214499 (NAMASTE), EPSRC-GB (GR/S81407/01) and EPSC-NSFC joint grant 10911130232/A0402. K. Y. W. acknowledges also the support of Chinese Academy of Sciences "100 talent program".

**Figure Captions :**

**Figure 1.** (a) optical micrograph of the Hall bar structure with bar width 4 µm and the length of the neighbour voltage contacts 20 µm. The lighter region on the left of the Hall bar has been etched to a depth of 10-15 nm. (b) Temperature dependence of the sheet resistance for the etched (open circles, right axis) and unetched (solid squares, left axis) regions of the Hall bar. (c) Temperature dependence of the Hall angle for the etched (open circles) and unetched (solid squares) regions of the Hall bar. (d) Temperature dependence of anisotropic magnetoresistance. Inset to (d): magnetoresistance for magnetic field perpendicular to plane (black,B//z) and magnetic field in-plane parallel (blue,B//x) and perpendicular to current (red,B//y). The schematic geometry is shown in the inset of (c).

**Figure 2.** PMOKM images of different magnetic configurations of the Hall bar channel: (a) 1 domain wall between the voltage probes; (b) 2 domain walls; (c) 2 domain walls; (d) 3 domain walls. The vertical line separates etched and unetched regions.

**Figure 3.** (a) Domain wall resistance for 1, 2 and 3 domain walls pinned in the unetched region of the Hall bar; (b) Domain wall resistance for domain wall at the etching step (stars) compared to the semiclassical calculation for W=0, 20 and 40nm (black, red and green line, respectively); (c) Average domain wall resistance in the unetched region (squares) compared to the semiclassical calculation; (d) Resistance-area product after subtracting extrinsic and AMR contributions, for the etching step (stars) and for the unetched region (squares).

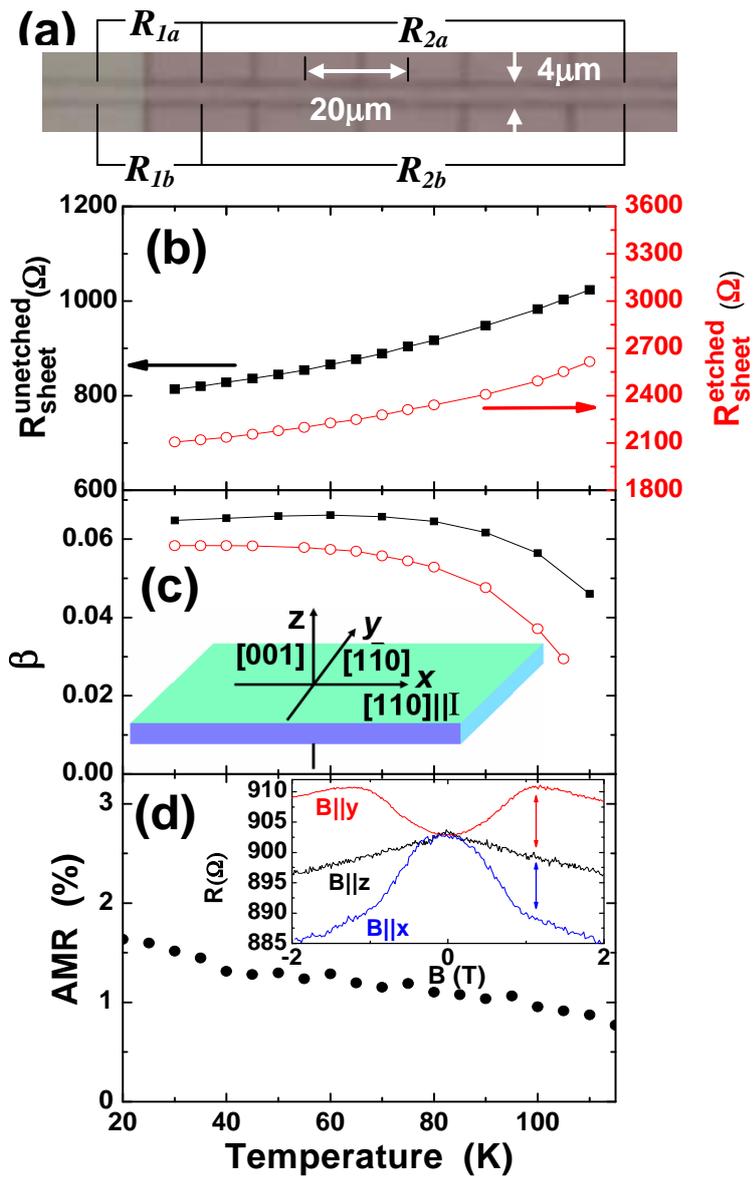

Figure 1   K. Y. Wang et al.

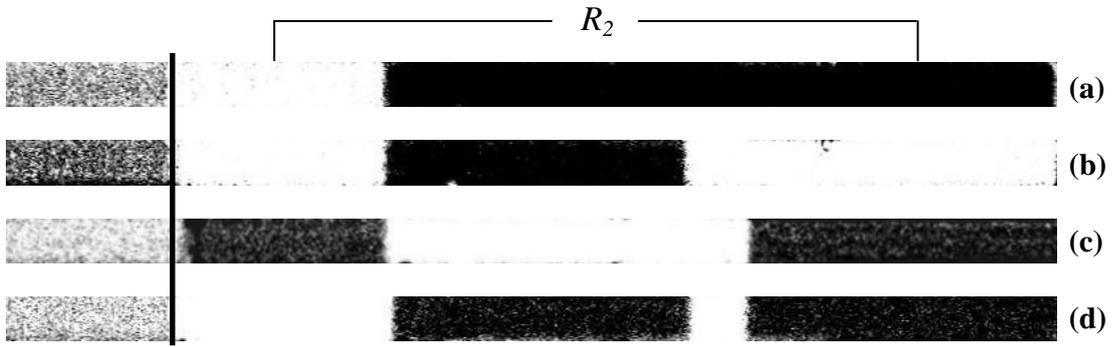

*Figure 2 K. Y. Wang et al.*

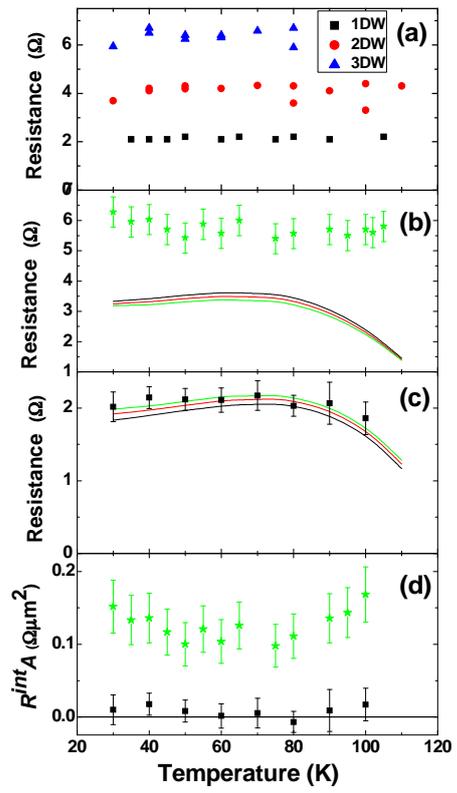

*Figure 3 K. Y. Wang et al.*